# Market Bubbles and Crashes

T. Kaizoji[1] and D. Sornette[2,3]

[1]Division of Social Sciences, International Christian University
3-10-2 Osawa, Mitaka, Tokyo 181-8585, Japan

[2]ETH Zurich, Department of Management, Technology and Economics,
Kreuzplatz 5, CH-8032 Zurich, Switzerland

[3]Swiss Finance Institute
c/o University of Geneva, 40 blvd. Du Pont d'Arve
CH 1211 Geneva 4, Switzerland

kaizoji@icu.ac.jp and dsornette@ethz.ch

**Abstract:** Episodes of market crashes have fascinated economists for centuries. Although many academics, practitioners and policy makers have studied questions related to collapsing asset price bubbles, there is little consensus yet about their causes and effects. This review and essay evaluates some of the hypotheses offered to explain the market crashes that often follow asset price bubbles. Starting from historical accounts and syntheses of past bubbles and crashes, we put the problem in perspective with respect to the development of the efficient market hypothesis. We then present the models based on heterogeneous agents and the limits to arbitrage that prevent rational agents from bursting bubbles before they inflate. Then, we explore another set of explanations of why rational traders would be led to actually profit from and surf on bubbles, by anticipating the behavior of noise traders or by realizing the difficulties in synchronizing their actions. We then end by discussing a complex system approach of social imitation leading to collective market regimes like herding and the phenomenon of bifurcation (or phase transition) that rationalize what crash can occur in unstable market regimes. The key insight is that diagnosing bubbles may be feasible when taking into account the positive feedback mechanisms that give rise to transient "super-exponential" price growth, the bubbles.

# I. Irrational Euphoria

The two acclaimed classic books, Galbraith's "The Great Crash 1929 (Galbraith, 1954) and Kindleberger's "Manias, Panics and Crash" (Kindleberger, 1978) provide the most commonly accepted explanation of the 1929 boom and crash. Galbraith argues that a bubble in the stock market was formed during the rapid economic growth of the 1920s. Both him and Kindleberger in his extensive historical compendium of financial excesses emphasize the irrational element -the mania- that induced the public to invest in the bull "over-heating" market. The rise in the stock market, according to Galbraith's account (1954 and 1988, pp. xii-xiii), depended on "the vested interest in euphoria [that] leads men and women, individuals and institutions to believe that all will be better, that they are meant to be richer and to dismiss as intellectually deficient what is in conflict with that conviction." This eagerness to buy stocks was then fueled by an expansion of credit in the form of brokers' loans that encouraged investors to become dangerously leveraged. In this respect, Shiller (2000) argues that the stock price increase was driven by irrational euphoria among individual investors, fed by an emphatic media, which maximized
TV ratings and catered to investor demand for pseudo-news.

Kindleberger (1978) summarizes his compilation of many historical bubbles as follows.
· The upswing usually starts with an opportunity - new markets, new technologies or some dramatic political change - and investors looking for good returns.
· It proceeds through the euphoria of rising prices, particularly of assets, while an expansion of credit inflates the bubble.
· In the manic phase, investors scramble to get out of money and into illiquid things such as stocks, commodities, real estate or tulip bulbs: 'a larger and larger group of people seeks to become rich without a real understanding of the processes involved'.
· Ultimately, the markets stop rising and people who have borrowed heavily find themselves overstretched. This is 'distress', which generates unexpected failures, followed by 'revulsion' or 'discredit'.
· The final phase is a self-feeding panic, where the bubble bursts. People of wealth and credit scramble to unload whatever they have bought at greater and greater losses, and cash becomes king.

While this makes for compelling reading, many questions remain unanswered. There is little consideration about how much fundamentals contributed to the bull market and what might have triggered the speculative mania. Galbraith (1954) cited margin buying, the formation of closed-end investment trusts, the transformation of financiers into celebrities, and other qualitative signs of euphoria to support his view. Recent evidence supports the concept of the growth of a social procyclical mood that promotes the attraction for investing in the stock markets by a larger and larger fraction of the population as the bubble grows (Roehner and Sornette, 2000).

Furthermore, Galbraith's and Kindleberger's accounts are vague about the causes of the market crash, believing that almost any event could have triggered irrational investors to sell towards the end of bubble, not really explaining the reason for the crash. Instead, they side-step the thorny question of the occurrence and timing of the crash by focusing on the inevitability of the bubble's collapse and suggest several factors that could have exploded public confidence and caused prices to plummet.

Furthermore, little has been done to identify the precise role of external events in provoking the collapse.

In the words of Shiller (1989), a crash is a time when "the investing public en masse capriciously changes its mind." But, as with the more rational theories, this explanation again leaves unanswered the question why such tremendous capricious changes in sentiment occur. Other studies have argued that even though fundamentals appeared high in 1929, Irving Fisher (1930), for example, argued throughout 1929 and 1930 that the high level of prices in 1929 reflected an expectation that future corporate cash flows would be very high. Fisher believed this expectation to be warranted after a decade of steadily increasing earnings and dividends, rapidly improving technologies, and monetary stability. In hindsight, it has become clear that even though fundamentals appeared high in 1929, the stock market rise was clearly excessive. A recent empirical study (De Long and Shleifer, 1991)) concludes that the stocks making up the S&P500 composite were priced at least 30 percent above fundamentals in late summer, 1929. Eugene N. White (2006) suggests that the 1929 boom can not be readily explained by fundamentals, represented by expected dividend growth or changes in the equity premium.

While Galbraith's and Kindleberger's classical views have been most often cited by the mass media, they had received little scholarly attention. Since the 1960s, in parallel with the emergence of the efficient-market hypothesis, their position has lost ground among economists and especially financial economists.

**II. Efficient Market Hypothesis**

The hypothesis that actual prices reflect fundamental values is the Efficient Markets Hypothesis (EMH). Market efficiency is of paramount importance to financial economists because it lays the foundations of economics and finance. And, therefore, it is no surprise that over the last several years, a large volume of empirical work has attempted to justify a variety of ways in which markets are efficient. Market efficiency assumes two things: no limited arbitrage as described above and agents' rationality. The agents' rationality means that agents update their beliefs correctly when they receive new information, and the subjective distribution they use to forecast future realizations of asset prices and returns is indeed the distribution from which those realizations are drawn.

In an efficient market, an asset price equals its "fundamental value". This is defined as the discounted sum of expected future cash flows where, in forming expectations, investors correctly process all available information. Therefore, in an efficient market, there is "no free lunch": no investment strategy can earn excess risk-adjusted average returns, or average returns greater than are warranted for its risk. Proponents of the efficient markets hypothesis (Friedman, 1953; Fama, 1965) argue that rational speculative activity would not only eliminate riskless arbitrage opportunities. Fama (1965, p.38) states that, if there are many sophisticated traders in the market, they may cause these bubbles to burst before they have a chance to really get under way.

To illustrate this argument, suppose that the fundamental value of a share of Yahoo! is $30. Imagine that a group of irrational traders becomes excessively pessimistic about

Yahoo!'s future prospects and through its selling, pushes the price to $20. Defenders of the EMH argue that rational traders, sensing an attractive opportunity, will buy the security at its bargain price and at the same time, hedge their bet by shorting a "substitute" security, such as Google, that has similar cash flows to Yahoo! in future states of the world. The buying pressure on Yahoo! Shares will then bring their price back to fundamental value of $30. Therefore, in an efficient market, bubbles do not exist.

However, after years of effort, it has become clear that some basic empirical facts about the stock markets cannot be understood in this framework (West, 1988). The efficient markets hypothesis lost ground entirely after the burst of the internet bubble in 2000, providing one of the recent most striking episodes of anomalous price behavior and volatility in the most well developed capital markets of the world. The movement of Internet stock prices during the late 1990s was extraordinary in many respects. The Internet sector earned over 1000 percent returns on its public equity in the two year period from early 1998 through February 2000. The valuations of these stocks began to collapse shortly thereafter and by the end of the same year, they had returned to pre-1998 levels, losing nearly 70% from the peak. The extraordinary returns of 1998 to Feb. 2000 had largely disappeared by the end of 2000. Although on February of 2000, the vast majority of internet-related companies had negative earnings, the Internet sector in U.S. was equal to 6% of the market capitalization of all U.S. public companies, and 20% of the publicly traded volume of the U.S. stock market (see, e.g., Ofek and Richardson, 2002; 2003)).

Ofek and Richardson (2003) used the financial data from 400 companies in the Internet-related sectors and analyzed how large their stock prices differed from their fundamental values estimated by using Miller and Modigliani's (1961) model for stock valuation (see French and Poterba, 1991)). Since almost all companies in the Internet sector had negative earnings, they estimated the (implied) price-to-earnings ratio (P/E) ratios which are derived from the revenue streams of these firms rather than their earnings that would be read from the 1999 financial data. Their results are striking. Almost 20% of the internet-related firms have P/E ratios in excess of 1500, while over 50% exceed 500, and the aggregate P/E ratio of the entire Internet sector is 605. Under the assumptions that the aggregate long-run P/E ratio is 20 on average (which is already on the large end member from an historical point of view), the Internet sector would have needed to generate 40.6% excess returns over a 10-year period to justify the P/E ratio of 605 implied in 2000. The vast majority of the implied P/Es are much too high relative to the P/Es usually obtained by firms. By almost any standard, this clearly represented "irrational" valuation levels. These and similar figures led many to believe that this set of stocks was in the midst of an asset price bubble.

### III. Heterogeneous Beliefs and Limit to Arbitrage

The collapsing Internet bubble have thrown new light on the old subject and raised the acute question of why have not rational investors moved into the market and driven the Internet stock prices back to their fundamental valuations?

Two conditions are in general invoked as being necessary for prices to deviate from fundamental value. First, there must be some degree of irrationality in the market. That is, investors' demand for stocks must be driven by something other than fundamentals, like overconfidence in the future. Second, even if a market has such investors, the general argument is that rational investors will drive prices back to fundamental value. For this not to happen, there needs to be some limits on arbitrage. Shleifer and Vishny (1997) provide a description for various limits of arbitrage. With respect to the equity market, clearly the most important impediment to arbitrage are short sales restrictions. Roughly 70% of mutual funds explicitly state (in SEC Form N-SAR) that they are not permitted to sell short (Almazan et al., 2004). Seventy-nine percent of equity mutual funds make no use of derivatives whatsoever (either futures or options), suggesting further that funds do not take synthetically short positions (Koski and Pontiff, 1999)). These figures indicate that the vast majority of funds never take short positions.

Recognizing that the world has limited arbitrage and a significant numbers of irrational investors, the finance literature has evolved to increasingly recognize the evidence of deviations from fundamental value. One important class of theories shows that there can be large movements in asset prices due to the combined effects of heterogeneous beliefs and short-sales constraints. The basic idea finds its root back to the original CAPM theories, in particular, Lintner (1969)'s model of asset prices with investors having heterogeneous beliefs. In his model, asset prices are a weighted average of beliefs about asset payoffs with the weights being determined by the investor's risk aversion and beliefs about asset price covariances. Lintner (1969), and many others after him, show that widely inflated prices can occur.

Many other asset pricing models in the spirit of Lintner (1969) have been proposed (See, for example, Miller (1977), Jarrow (1980), Harrison and Kreps (1978), Chen, Hong and Stein (2000, 2002), Scheinkman and Xiong (2003) and Duffie, Garleanu and Pedersen (2002)). In these models which assume heterogeneous beliefs and short sales restrictions, the asset prices are determined at equilibrium to the extent that they reflect the heterogeneous beliefs about payoffs, but short sales restrictions force the pessimistic investors out of the market, leaving only optimistic investors and thus inflated asset price levels. However, when short sales restrictions no longer bind investors, then prices fall back down. This provides a possible account of the bursting of the Internet bubble that developed in 1998-2000. As documented by Ofek and Richardson (2003) and Cochrane (2003), typically as much as 80% of Internet-related shares were locked up. This is due to the fact that many Internet companies had gone through recent initial public offerings (IPOs) and regulations impose that shares held by insiders and other pre-IPO equity holders cannot be traded for at least 6 months after the IPO date. The float of the Internet sector dramatically increased as the lockups of many of these stocks expired. The unlocking of literally hundreds of billions of dollars of shares in the Internet sector in Spring 2000 was the equivalent of removing short sales restrictions. And the collapse of Internet stock prices coincided with a dramatic expansion in the number of Internet companies publicly tradable shares. Among many, Hong, Scheinkman and Xiong (2006) model explicitly the relationship between the number of publicly tradable shares of an asset and the propensity for speculative bubbles to form. So far, the theoretical models based on agents with heterogeneous beliefs facing short sales restrictions are considered among the most convincing models to explain the burst of the Internet bubbles.

Another test of this hypothesis on the origin of the 2000 market crash is provided by the search for possible discrepancies between option and stock prices. Indeed, even though it is difficult for rational investors to borrow Internet stocks for short selling due to the locked-up's discussed above, they should have been able to construct equivalent synthetic short positions by purchasing puts and writing calls in the option market and either borrowing or lending cash, without the need for borrowing the stocks. The question is now transformed into finding some evidence for the use or absence of such strategy and why in the later case. One possible thread is that, if short selling through option positions was difficult or impractical, prices in the stock and options markets should decouple (Lamont and Thaler, 2003). Using a sample of closing bid and ask prices for 9026 option pairs for three days in February 2000 along with closing trade prices for the underlying equities, Ofek and Richardson (2003) find that 36% of the Internet stocks had put-call parity violations as compared to only 23.8% of the other stocks. One reason for put-call parity violations may be that short-sale restrictions prevent arbitrage from equilibrating option and stock prices. Hence, one interpretation of the finding that there are more put-call parity violations for Internet stocks is that short-sale constraints are more frequently binding for Internet stocks. Furthermore, Ofek, Richardson, and Whitelaw (2004) provide a comprehensive comparison of the prices of stocks and options, using closing options quotes and closing trades on the underlying stock for July 1999 through November 2001. They find that there are large differences between the synthetic stock price and the actual stock price, which implies the presence of apparent arbitrage opportunities involving selling actual shares and buying synthetic shares. They interpret their findings as evidence that short-sale constraints provide meaningful limits to arbitrage that can allow prices of identical assets to diverge.

It would thus seem that the issue of the origin of the 2000 crash is settled. However, Battalio and Schultz (2006) arrive at the opposite conclusion using proprietary intraday option trade and quote data generated in the days surrounding the collapse of the Internet bubble. They find that the general public could cheaply short synthetically using options, and this information could have been transmitted to the stock market, in line with the absence of evidence that synthetic stock prices diverged from actual stock prices. The difference between Ofek and Richardson (2003) and Ofek, Richardson, and Whitelaw (2004) on the one hand and Battalio and Schultz (2006) on the other hand is that the former used closing option quotes and last stock trade prices from the OptionMetrics Ivy database. As pointed out by Battalio and Schultz (2006), OptionMetrics matches closing stock trades that occurred no later than 4:00pm, and perhaps much earlier, with closing option quotes posted at 4:02pm. Furthermore, option market makers that post closing quotes on day t are not required to trade at those quotes on day t+1. Likewise, dealers and specialists in the underlying stocks have no obligation to execute incoming orders at the price of the most recent transaction. Hence, closing option quotes and closing stock prices obtained from the OptionMetrics database do not represent contemporaneous prices at which investors could have simultaneously traded. To address this problem, Battalio and Schultz (2006) uses a unique set of intraday option price data. They first ensure that the synthetic and actual stock prices that they compare are synchronous, and then, they discard quotes that, according to exchange rules, are only indicative of the prices at which liquidity demanders could have traded. They find that almost all of the remaining apparent put-call parity violations disappear when they discard locked or crossed quotes and quotes from fast options markets. In other words, the apparent arbitrage opportunities almost always arise from quotes upon which investors could not actually trade. Battalio and Schultz (2006) conclude that short-sale constraints were not responsible for the high prices of Internet stocks at the peak of the bubble, that small investors could have sold short synthetically using options, and this information would have been transmitted to the stock market. The fact that investors did not take advantage of these opportunities to profit from overpriced Internet stocks suggests that the overpricing was not as obvious then as it is now with the benefit of hindsight. Schultz (2008) provides additional evidence that contemporaneous lockup expirations and equity offerings do not explain the collapse of Internet stocks because the stocks that were restricted to a fixed supply of shares by lockup provisions actually performed worse than stocks with an increasing supply of shares. This shows that current explanations for the collapse of Internet stocks are incomplete.

**IV. Ridding Bubbles**

What is the origin of bubbles? In a nutshell, speculative bubbles are caused by "precipitating factors" that change public opinion about markets or that have an immediate impact on demand, and by "amplification mechanisms" that take the form of price-to-price feedback, as stressed by Shiller (2000). Consider the example of a house bubble. A number of fundamental factors can influence price movements in housing markets. On the demand side, demographics, income growth, employment growth, changes in financing mechanisms or interest rates, as well as changes in location characteristics such as accessibility, schools, or crime, to name a few, have been shown to have effects. On the supply side, attention has been paid to construction costs, the age of the housing stock, and the industrial organization of the housing market. The elasticity of supply has been shown to be a critical factor in the cyclical behavior of home prices. The cyclical process that we observed in the 1980s in those cities experiencing boom-and-bust cycles was caused by the general economic expansion, best proxied by employment gains, which drove demand up. In the short run, those increases in demand encountered an inelastic supply of housing and developable land, inventories of for-sale properties shrank, and vacancy declined. As a consequence, prices accelerated. This provided an amplification mechanism as it led buyers to anticipate further gains, and the bubble was born. Once prices overshoot or supply catches up, inventories begin to rise, time on the market increases, vacancy rises, and price increases slow down, eventually encountering downward stickiness. The predominant story about home prices is always the prices themselves (Shiller, 2000; Sornette, 2003); the feedback from initial price increases to further price increases is a mechanism that amplifies the effects of the precipitating factors. If prices are going up rapidly, there is much word-of-mouth communication, a hallmark of a bubble. The word of mouth can spread optimistic stories and thus help cause an overreaction to other stories, such as ones about employment. The amplification can also work on the downside as well.

Hedge funds are among the most sophisticated investors, probably closer to the ideal of "rational arbitrageurs" than any other class of investors. It is therefore particularly telling that successful hedge fund managers have been repeatedly reported to ride rather than attack bubbles, suggesting the existence of mechanisms that entice rational investors to surf bubbles rather than attempt to arbitrage them. However, the evidence may be not that strong, and could even be circular, since only successful hedge-fund managers would survive a given 2-5 years period, opening the possibility that the mentioned evidence could result in large part from a survival bias (Brown et al., 1992; Grinblatt and Titman, 1992). Keeping this in mind, we now discuss two classes of models, which attempt to justify why sophisticated "rational" traders would be willing to ride bubbles. These models share a common theme: rational investors try to ride bubbles, and the incentive to ride the bubble stems from predictable "sentiment": anticipation of continuing bubble growth in Abreu and Brunnermeier (2003) and predictable feedback trader demand in De Long et al. (1990b). An important implication of these theories is that rational investors should be able to reap gains from riding a bubble at the expense of less sophisticated investors.

**IV-1. Positive Feedback Trading by Noise Traders**

The term "noise traders" was introduced first by Kyle (1985) and Black (1986) to describe irrational investors. Thereafter, many scholars exploited this concept to extend the standard models by introducing the simplest possible heterogeneity in terms of two interacting populations of rational and irrational agents. One can say that the one-representative-agent theory is being progressively replaced by a two-representative-agents theory, analogously to the progress from the one-body to the two-body problems in Physics.

De Long, Shleifer, Summers and Waldmann (1990b) introduced a model of market bubbles and crashes which exploits this idea of the possible role of noise traders in the development of bubbles, as a possible mechanism for why asset prices may deviate from the fundamentals over rather long time periods. Their inspiration came from the observation of successful investors such as George Soros, who reveal that they often exploit naive investors following positive feedback strategies or momentum investment strategies. Positive feedback investors are those who buy securities when prices rise and sell when prices fall. In the words of Jegadeesh and Titman (1993), positive feedback investors are buying winners and selling losers. In George Soros (1987) description of his own investment strategy, he stresses that the key to his success was not to counter the irrational wave of enthusiasm which appears in financial markets, but rather to *ride this wave* for a while and sell out much later. The model of De Long et al. (1990b) assumes that, when rational speculators receive good news and trade on this news, they recognize that the initial price increase will stimulate buying by noise traders who will follow positive feedback trading strategies with a delay. In anticipation of these purchases, rational speculators buy more today, and so drive prices up today higher than fundamental news warrants. Tomorrow, noise traders buy in response to today's price increase and so keep prices above fundamentals. The key point is that trading between rational arbitrageurs and positive feedback traders gives rise to bubble-like price patterns. In their model, rational speculators destabilizes prices because their trading triggers positive feedback trading by other investors. Positive feedback trading reinforced by arbitrageurs' jumping on the bandwagon leads to a positive auto-correlation of returns at short horizons. Eventually selling out or going short by rational speculators will pull the prices back to fundamentals, entailing a negative autocorrelation of returns at longer horizons. In summary, De Long et al. (1990b)'s model suggests the coexistence of intermediate horizon momentum and long horizon reversals in stock returns.

Their work was followed by a number of behavioral models based on the idea that trend chasing by one class of agents produces momentum in stock prices (Barberis, Shleifer, and Vishny, 1998; Daniel, Hirshleifer, and Subrahmanyam, 1998; Hong and Stein, 1999). The most influential empirical evidence on momentum strategies came from the work of Jegadeesh and Titman (1993, 2001), which established that stock returns exhibit momentum behavior at intermediate horizons. Strategies which buy stocks that have performed well in the past and sell stocks that have performed poorly in the past generate significant positive returns over 3- to 12- month holding periods. De Bondt and Thaler (1985) documented long-term reversals in stock returns. Stocks that perform poorly in the past perform better over the next 3 to 5 years than stocks that perform well in the past. These findings present a serious challenge to the view that markets are semi strong-form efficient.

In practice, do investors engage in momentum trading? A growing number of empirical studies address momentum trading by investors, with somewhat conflicting results. Lakonishok et al. (1992) analyzed the quarterly holdings of a sample of pension funds and found little evidence of momentum trading. Grinblatt, Titman, and Wermers (1995) examined the quarterly holdings of 274 mutual funds and found that 77 percent of the funds in their sample engage in momentum trading (see also Wermers (1999)). Nofsinger and Sias (1999) examined total institutional holdings of individual stocks and foundd evidence of intra-period momentum trading. Using a different sample, Gompers and Metrick (2001) investigated the relation between institutional holdings and lagged returns and concluded that once they controlled for firm size, there was no evidence of momentum trading. Griffin, Harris, and Topaloglu (2003) reported that, on a daily and intraday basis, institutional investors engaged in trend-chasing in Nasdaq 100 stocks. Finally, Badrinath and Wahal (2004) documented the equity trading practices of approximately 1,200 institutions from the third quarter of 1987 through the third quarter of 1995. They decomposed trading by institutions into (i) the initiation of new positions (entry), (ii) the termination of previous positions (exit), and (iii) the adjustments to ongoing holdings. Institutions were found to act as momentum traders when they enter stocks but as contrarian traders when they exit or make adjustments to ongoing holdings. Badrinath and Wahal (2004) found significant differences in trading practices among different types of institutions. These studies are limited in their ability to capture the full range of trading practices, in part because they focus almost exclusively on the behavior of institutional investors. In summary, many experimental studies and surveys suggest that positive feedback trading exists in greater or lesser degrees.

**IV-2. Synchronization Failures among Rational Traders**

Abreu and Brunnermeier (2003) propose a completely different mechanism justifying why rational traders ride rather than arbitrage bubbles. They consider a market where arbitrageurs face synchronization risk and, as a consequence, delay usage of arbitrage opportunities. Rational arbitrageurs are supposed to know that the market will eventually collapse. They know that the bubble will burst as soon as a sufficient number of (rational) traders will sell out. However, the dispersion of rational arbitrageurs' opinions on market timing and the consequent uncertainty on the synchronization of their sell-off are delaying this collapse, allowing the bubble to grow. In this framework, bubbles persist in the short and intermediate term because short sellers face synchronization risk, that is, uncertainty regarding the timing of the correction. As a result, arbitrageurs who conclude that othe arbitrageurs are yet unlikely to trade against the bubble find it optimal to ride the still-growing bubble for a while.

Like other institutional investors, hedge funds with large holdings in U.S. equities have to report their quarterly equity positions to the SEC on Form 13F. Brunnermeier and Nagel (2004) extracted hedge fund holdings from these data, including those of well-known managers such as Soros, Tiger, Tudor, and others in the period from 1998 to 2000. They found that, over the sample period 1998 to 2000, hedge fund portfolios were heavily tilted toward highly priced technology stocks. The proportion of their overall stock holdings devoted to this segment was higher than the corresponding weight of technology stocks in the market portfolio. In addition, the hedge funds in their sample skillfully anticipated price peaks of individual technology stocks. On a stock-by-stock basis, hedge-funds started to cut back their holdings before prices collapsed, switching to technology stocks that still experienced rising prices. As a result, hedge fund managers captured the upturn, but avoided much of the downturn. This is reflected in the fact that hedge funds earned substantial excess returns in the technology segment of the Nasdaq.

## V. Complex system theory of bubbles and crashes

Bhattacharya and Yu (2008) provide a summary of recent efforts to expand on the above concepts, in particular to address the two main questions of (i) the cause(s) of bubbles and crashes and (ii) the possibility to diagnose them ex-ante. Many financial economists recognize that positive feedbacks and in particular herding is a key factor for the growth of bubbles. Herding can result from a variety of mechanisms, such as anticipation by rational investors of noise traders' strategies (De Long et al., 1990b), agency costs and monetary incentives given to competing fund managers (Dass, Massa and Patgiri, 2008) sometimes leading to the extreme Ponzi schemes (Dimitriadi, 2004), rational imitation in the presence of uncertainty (Roehner and Sornette, 2000), and social imitation. Discussing social imitation is often considered off-stream among financial economists but warrants some scrutiny, given its pervasive presence in human affairs. On the question of the ex-ante detection of bubbles, Gurkaynak (2008) summarizes the dismal state of the econometric approach, stating that the "econometric detection of asset price bubbles cannot be achieved with a satisfactory degree of certainty. For each paper that finds evidence of bubbles, there is another one that fits the data equally well without allowing for a bubble. We are still unable to distinguish bubbles from time-varying or regime-switching fundamentals, while many small sample econometrics problems of bubble tests remain unresolved." The following discusses an arguably off-stream approach which, by using concepts and tools from the theory of complex systems and statistical physics, suggests that ex-ante diagnostic and partial predictability might be possible (see the review in (Sornette, 2003))

### V-1 Social Imitation, collective phenomena, bifurcations and phase transitions

One of the most robust characteristics of humans, which has probably the most visible imprint in our social affairs, is imitation and herding. Imitation, in obvious or subtle forms, is a pervasive activity of humans. Imitation has been documented in psychology and in neuro-sciences as one of the most elaborated cognitive process, requiring a developed cortex and sophisticated processing abilities. In short, we learn what we know and how to adapt mostly by imitation all along our life. It seems that imitation has evolved as an evolutionary advantageous trait, and may even have promoted the development of our anomalously large brain compared with other mammals (Dunbar, 1998). It is actually rational to imitate when lacking sufficient time, energy and information to take a decision based only on private information and processing, that is ... most of the time.

Market behavior is the aggregation of the individual behavior of the many investors participating in it. In an economy of traders with completely rational expectations and the same information sets, no bubbles are possible (Tirole, 1982). Rational bubbles can however occur in infinite horizon models (Blanchard and Watson, 1982), with dynamics of growth and collapse driven by noise traders (Johansen, Sornette and Ledoit, 1999; Johansen, Ledoit and Sornette, 2000). But the key issue is to understand by what detailed mechanism the aggregation of many individual behaviors can give rise to bubbles and crashes. Modeling social imitation and social interactions requires using approaches, little known to financial economists, that address the fundamental question of how global behaviors can emerge at the macroscopic level. This extends the representative agent approach, of course, but goes also well beyond the introduction of heterogeneous agents. A key insight from statistical physics and complex system theory is that systems with a large number of interacting agents, open to their environment, self-organize their internal structure and their dynamics with novel and sometimes surprising "emergent" out-of-equilibrium properties. A central property of a complex system is the possible occurrence and co-existence of many large-scale collective behaviors with a very rich structure, resulting from the repeated non-linear interactions among its constituents.

How can this help address the question of what is/are the cause(s) of bubbles and crashes? The crucial insight is that a system, made of competing investors subjected to the myriad of influences, both exogenous news and endogenous interactions and reflexivity, can develop into endogenously self-organized self-reinforcing regimes, that would qualify as bubbles, and that crashes occur as a global self-organized transition. Mathematicians refer to this behavior as a "bifurcation" or more specifically as a "catastrophe" (Thom, 1989). Physicists call these phenomena "phase transitions" (Stanley, 1987). The implication of modeling a market crash as a bifurcation is to solve the question of what makes a crash: in the framework of bifurcation theory (or phase transitions), sudden shifts in behavior arise from small changes in circumstances, with qualitative changes in the nature of the solutions which can occur abruptly when the parameters change smoothly. A minor change of circumstances, of interaction strength, or heterogeneity may lead to a sudden and dramatic change, such as during an earthquake and a financial crash.

Most approaches to explaining crashes search for possible mechanisms or effects that operate at very short time scales (hours, days, or weeks at most). According to the "bifurcation" approach, the underlying cause of the crash should be found in the preceding months and years, in the progressively increasing build-up of market cooperativity, or effective interactions between investors, often translated into accelerating ascent of the market price (the bubble). According to this "critical" point of view, the specific manner by which prices collapsed is not the most important problem: a crash occurs because the market has entered an unstable phase and any small disturbance or process may reveal the existence of the instability.

**V-2 The class of Ising models of Social Imitation and phase transitions**

Perhaps the simplest and historically most important model describing how the aggregation of many individual behaviors can give rise to macroscopic out-of-equilibrium dynamics such as bubbles, with bifurcations in the organization of social systems due to slight changes in the interactions is the Ising model (Callen and Shapero, 1974; Montrol and Badger, 1974). In particular, Orléan (1989, 1995) has captured the paradox of combining rational and imitative behavior under the name "mimetic rationality," by developing models of mimetic contagion of investors in the stock markets, which are based on irreversible processes of opinion forming. Roehner and Sornette (2000) among others showed that the dynamical updating rules of the Ising model are obtained in a natural way as the optimal strategy of rational traders with limited information who have the possibility to make up for their lack of information via information exchange with other agents within their social network. The Ising model is one of the simplest models describing the competition between the ordering force of imitation or contagion and the disordering impact of private information or idiosyncratic noise (see (McCoy and Wu, 1973) for a technical review).

Starting with a framework suggested by Blume (1993; 1995), Brock (1993), Durlauf (1991; 1993;1997; 1999), and Phan et al. (2004) summarize the formalism starting with different implementation of the agents' decision processes whose aggregation is inspired from statistical mechanics to account for social influence in individual decisions. Lux and Marchesi (1999), Brock and Hommes (1999), Taizoji (2000) and Kirman and Teyssiere (2002) have also developed related models in which agents' successful forecasts reinforce the forecasts. Such models have been found to generate swings in opinions, regime changes and long memory. An essential feature of these models is that agents are wrong for some of the time, but whenever they are in the majority they are essentially right. Thus, they are not systematically irrational (Kirman, 1997). Sornette and Zhou (2006) show how Bayesian learning added to the Ising model framework reproduces the stylized facts of financial markets. Harras and Sornette (2008) show how over-learning from lucky runs of random news in the presence of social imitation may lead to endogenous bubbles and crashes.

These models allow one to combine the questions on the cause of both bubbles and crashes, as resulting from the collective emergence of herding via self-reinforcing imitation and social interactions, which are then susceptible to phase transitions or bifurcations occurring under minor changes in the control parameters. Hence, the difficulty in answering the question of "what causes a bubble and a crash" may in this context be attributed to this distinctive attribute of a dynamical out-of-equilibrium system to exhibit bifurcation behavior in its dynamics. The line of thought has been pursued by one of us and his co-authors, to propose a novel operational diagnostic of bubbles.

**V-3 Bubble as super-exponential price growth, diagnostic and prediction**

Bubbles are often defined as exponentially explosive prices, which are followed by a sudden collapse. As summarized for instance by Gurkaynak (2008), the problem with this definition is that any exponentially growing price regime, that one would call a bubble, can be also rationalized by a fundamental valuation model. This is related to the problem that the fundamental price is not directly observable, giving no strong anchor to understand observed prices. This was exemplified during the last Internet bubble by fundamental pricing models, which incorporated real options in the fundamental valuation, justifying basically any price. Mauboussin and Hiler (1999) were among the most vocal proponents of the proposition offered close to the peak of the Internet bubble that culminated in 2000, that better business models, the network effect, first-to-scale advantages, and real options effect could account rationally for the high prices of dot-com and other New Economy companies. These interesting views expounded in early 1999 were in synchrony with the bull market of 1999 and preceding years. They participated in the general optimistic view and added to the strength of the herd. Later, after the collapse of the bubble, these explanations seemed less attractive. This did not escape U.S. Federal Reserve chairman Alan Greenspan (1997), who said: "Is it possible that there is something fundamentally new about this current period that would warrant such complacency? Yes, it is possible. Markets may have become more efficient, competition is more global, and information technology has doubtless enhanced the stability of business operations. But, regrettably, history is strewn with visions of such new eras that, in the end, have proven to be a mirage. In short, history counsels caution." In this vein, the buzzword "new economy" so much used in the late 1990s was also hot in the 1960s during the "tronic boom" also followed by a market crash, and during the bubble of the late 1920's before the Oct. 1929 crash. In this later case, the "new" economy was referring to firms in the utility sector. It is remarkable how traders do not learn the lessons of their predecessors!

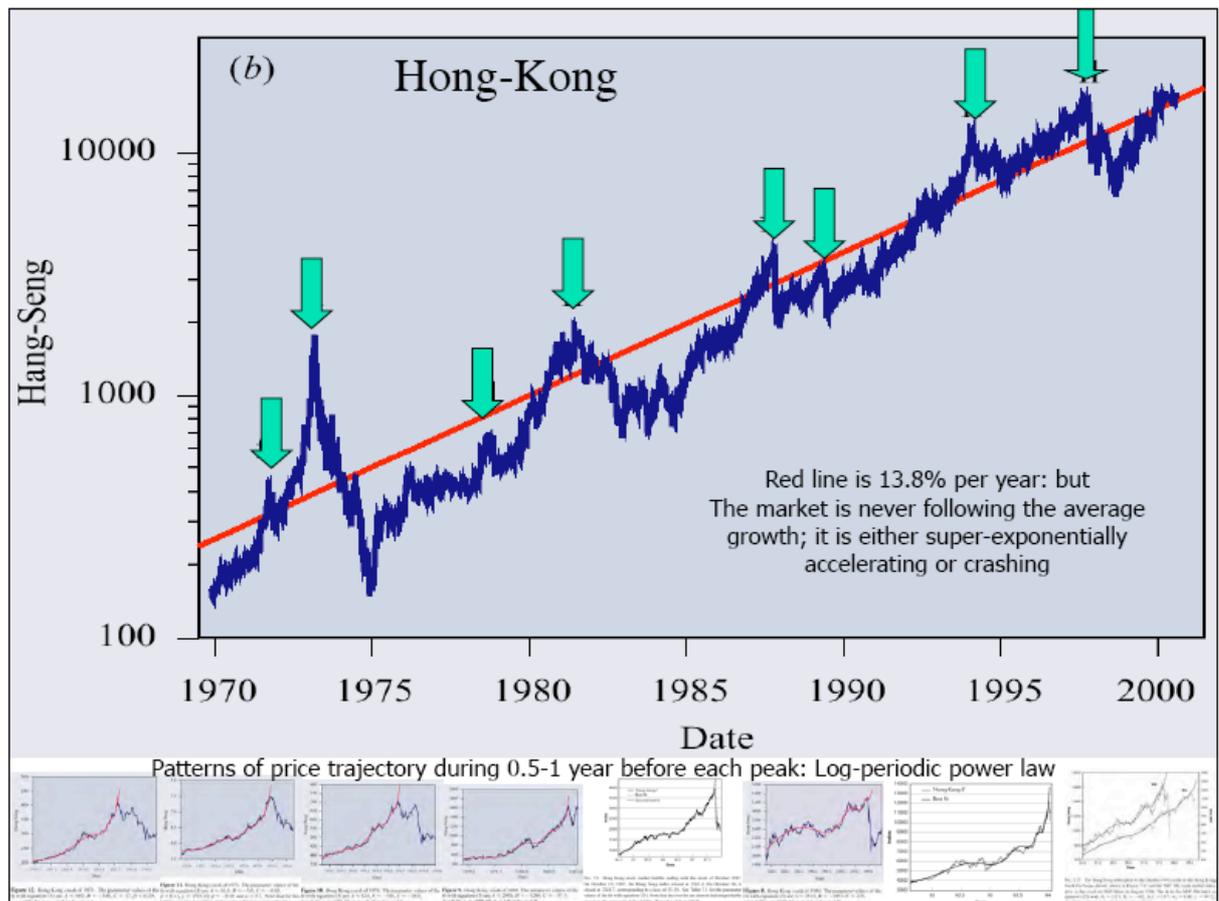

Fig.1: Trajectory of the Hong-Kong Hang Seng index from 1970 to 2000. The vertical log-scale together with the linear time scale allows one to qualify an exponential growth with constant growth rate as a straight line. This is indeed the long-term behavior of this market, as shown by the best linear fit represented by the solid straight line, corresponding to an average constant growth rate of 13.8% per year. The 8 arrows point to 8 local maxima that were followed by a drop of the index of more than 15% in less than three weeks (a possible definition of a crash). The 8 small panels at the bottom show the upward curvature of the log-price trajectory preceding each of these local maxima, which diagnose an unsustainable bubble regime, which culminates at the peak before crashing.

A better model derives from the mechanism of positive feedbacks discussed above, which gives rise generically to faster-than-exponential growth of price (coined hereafter "super-exponential") (Sornette and Andersen, 2002; Sornette, Takayasu and Zhou, 2003). This is nicely illustrated by the price trajectory of the Hong-Kong Hang Seng index from 1970 to 2000 shown in figure 1. The Hong Kong financial market is repeatedly rated as providing one of the most pro-economic, pro-entrepreneurship and free market-friendly environment in the world, and thus provides a textbook example of the behavior of weakly regulated liquid and striving financial markets. In figure 1, the logarithm of the index price $p(t)$ is plotted as a function of the time (in linear scale), so that an upward trending straight line qualifies as exponential growth with a constant growth rate: the straight solid line corresponds indeed to an approximately constant growth rate of the Hang Seng index equal to 13.8% per year. However, the most striking feature of figure 1 is not this average behavior, but the obvious fact that the real market is never following and abiding to a constant growth rate. One can observe a succession of price run-ups characterized by growth rates…growing themselves: this is reflected visually in figure 1 by transient regimes characterized by strong upward curvature of the price trajectory. Such an upward curvature in a linear-log plot is a first visual diagnostic of a faster than exponential growth (which of course needs to be confirmed by rigorous statistical testing).

Such super-exponential price trajectories are expected from the models of section V-2. An exponential growing price is characterized by a constant expected growth rate. The geometric random walk is the standard stochastic price model embodying this class of behaviors. A super-exponential growing price is such that the growth rate grows itself as a result of positive feedbacks of price, momentum and other characteristics on the growth rate (Sornette and Andersen, 2002). As a consequence of the acceleration, the mathematical models generalizing the geometric random walk exhibit so-called finite-time singularities. In other words, the resulting processes are not defined for all times: the dynamics has to end after a finite life and to transform into something else. This captures well the transient nature of bubbles, and the fact that the crashes ending the bubbles are often the antechambers to different market regimes. Mathematically, the simplest formula to represent a super-exponential behavior is a power law with a finite-time singularity at some finite future time $t_c$, beyond which another regime emerges:

(1)     $ln[E[p(t)]] = A + B (t_c-t)^m$,     where $B<0$,  $0<m<1$ .

Here $t_c$ is the theoretical critical time corresponding to the end of the transient run-up (end of the bubble). Such transient faster-than-exponential growth of the expected price $E[p(t)]$ is the definition of a bubble according to Sornette and co-workers. It has the major advantage of avoiding the conundrum of distinguishing between exponentially growing fundamental price and exponentially growing bubble price, which is a problem permeating most of the previous statistical tests developed to identify bubbles. The conditions $B<0$ and $0<m<1$ ensure the super-exponential acceleration of the price, together with the condition that the price remains finite even at $t_c$. Stronger singularities can appear for $m<0$.

Such a mathematical expression (1) is obtained from models that capture the effect of a positive feedback mechanism. Let us illustrate it with the simplest example. Starting with a standard proportional growth process $dp/dt = r p$ (omitting for the sake of pedagogy the stochastic component), where $r$ is the growth rate, let us assume that r is itself an increasing function of the price $p$, as a result of the positive feedback of the price on the future returns. For illustration, let us assume that $r$ is simply proportional to $p$ ($r=c p$, where $c$ is a constant), so that the proportional growth equation become $dp/dt = c p^2$. The solution of this equation is of the form (1) where $ln[p(t)]$ is replaced by p(t), with $m=-1$ and $A=0$ corresponding to a divergence of $p(t)$ at $t_c$. Many systems exhibit similar transient super-exponential growth regimes, which are described mathematically by power law growth with an ultimate finite-time singular behavior: planet formation in solar systems by runaway accretion of planetesimals, Euler equation of inviscid fluids, general relativity coupled to a mass field leading to formation of black holes in finite time, Zakharov equation of beam-driven Langmuir turbulence in plasma, rupture and material failures, nucleation of earthquakes modeled with the slip-and-velocity weakening Ruina-Dieterich friction law, models of micro-organisms interacting through chemotaxis aggregating to form fruiting bodies, Mullins-Sekerka surface instability, jets from a singular surface, fluid drop snap-off, the Euler rotating disk, and so on. Such mathematical equations can actually provide an accurate description of the transient dynamics, not too close to the mathematical singularity where new mechanisms come into play. The singularity at $t_c$ mainly signals a change of regime. In the present context, $t_c$ is the end of the bubble and the beginning of a new market phase, possible a crash or a different regime.

Such an approach may be thought of at first sight to be inadequate or too naive to capture the intrinsic stochastic nature of financial prices, whose null hypothesis is the geometric random walk model (Malkiel, 1990). However, it is possible to generalize this simple deterministic model to incorporate nonlinear positive feedback on the stochastic Black-Scholes model, leading to the concept of stochastic finite-time singularities (Sornette and Andersen, 2002; Fogedby and Poukaradzez, 2002; Fogedby, 2003; Andersen and Sornette, 2004). Still much work needs to be done on this theoretical aspect.

Coming back to figure 1, one can also notice that each burst of super-exponential price growth is followed by a crash, here defined for the eight arrowed cases as a correction of more than 15% in less than 3 weeks. These examples suggest that the non-sustainable super-exponential price growths announced a "tipping point" followed by a price disruption, i.e., a crash. The Hong-Kong Hang Seng index provides arguably one of the best textbook example of a free market in which bubbles and crashes occur repeatedly: the average exponential growth of the index is punctuated by a succession of bubbles and crashes, which seem to be the norm rather than the exception.

More sophisticated models than (1) have been proposed to take into account the interplay between technical trading and herding (positive feedback) versus fundamental valuation investments (negative mean-reverting feedback). Accounting for the presence of inertia between information gathering and analysis on the one hand and investment implementation on the other hand (Ide and Sornette, 2002) or between trend followers and value investing (Farmer, 2002), the resulting price dynamics develop second-order oscillatory terms and boom-bust cycles. Value investing does not necessarily cause prices to track value. Trend following may cause short-term trend in prices, but also cause longer-term oscillations. The simplest model generalizing (1) and including these ingredients is the so-called log-periodic power law (LPPL) model (see Sornette (2003) and references therein). Formally, some of the corresponding formulas can be obtained by considering that the exponent $m$ is a complex number with an imaginary part, where the imaginary part expresses the existence of a preferred scaling ratio $g$ describing how the continuous scale invariance of the power law (1) is partially broken into a discrete scale invariance (Sornette, 1998). The LPPL structure may also reflect the discrete hierarchical organization of networks of traders, from the individual to trading floors, to branches, to banks, to currency blocks. More generally, it may reveal the ubiquitous hierarchical organization of social networks recently reported (Zhou et al., 2005) to be associated with the social brain hypothesis (Dunbar, 1998).

Examples of calibrations of financial bubbles with one implementation of the LPPL model are the 8 super-exponential regimes discussed above in figure 1: the 8 small insets at the bottom of figure 1 show the LPPL calibration on the Hang Seng index. Preliminary tests reported (Sornette, 2003) suggest that the LPPL model provides a good starting point to detect bubbles and forecast their most probable end. Rational expectation models of bubbles a la Blanchard and Watson implementing the LPPL model (Johansen et al., 1999, 2000) have shown that the end of the bubble is not necessarily accompanied by a crash, but it is indeed the time where a crash is the most probable. But crashes can occur before (with smaller probability) or not at all. That is, a bubble can land smoothly, approximately one-third of the time, according to preliminary investigations (Johansen and Sornette, 2004). Therefore, only probabilistic forecasts can be developed. Probability forecasts are indeed valuable and commonly used in daily life, such as in weather forecast.

In a series of empirical papers, the second author and his collaborators have used this concept to test empirically for bubbles. Johansen and Sornette (2004) provide perhaps the most inclusive series of tests of this approach. First, they identify the most extreme cumulative losses (drawdowns) in a variety of asset classes, markets and epochs, and show that they belong to a probability density distribution, which is distinct from the distribution of 99% of the smaller drawdowns (the more "normal" market regime). These drawdowns can thus be called "outliers" or "kings". Second, they show that, for two-third of these extreme drawdowns, the market prices followed a super-exponential behavior prior to their occurrences, as characterized by a calibration of the power law with a finite-time singularity.

This approach provides a systematic approach to diagnose for bubbles ex-ante, as shown in a series of real-life tests (Sornette and Zhou, 2006; Sornette, Woodard and Zhou, 2008; Zhou and Sornette, 2003; 2006, 2007; 2008). While this approach has enjoyed a large visibility in the professional financial community around the world (banks, mutual funds, hedge-funds, investment houses, and so on), it has not yet received the attention from the academic financial community that it perhaps deserves given the stakes. This is probably due to several factors, which include: (i) the origin of the hypothesis coming from analogies with complex critical systems in Physics and the theory of complex systems, which constitutes a well-known obstacle to climb the ivory towers of standard financial economics; (ii) the non-standard (from an econometric view point) formulation of the statistical tests performed until present (in this respect, see the attempts in terms of a Bayesian analysis of LPPL precursors (Chang and Feigenbaum, 2006) to focus on the time series of returns instead of prices, and of regime-switching model of LPPL (Chang and Feigenbaum, 2007)), (iii) the non-standard expression of some of the mathematical models underpinning the hypothesis; and (iv) perhaps an implicit general belief in academia that forecasting financial instabilities is inherently impossible.